\def\be{\begin{equation}}
\def\ee{\end{equation}}
\def\ba{\begin{array}}
\def\ea{\end{array}}

\documentclass[aps,amsmath,amssymb,amsfonts,showpacs,twocolumn,pra]{revtex4}
\usepackage{graphicx}
\usepackage{caption2}
\usepackage{epstopdf}
\def\qed{\leavevmode\unskip\penalty9999 \hbox{}\nobreak\hfill
     \quad\hbox{\leavevmode  \hbox to.77778em{%
               \hfil\vrule   \vbox to.675em%
               {\hrule width.6em\vfil\hrule}\vrule\hfil}}
     \par\vskip3pt}

\newtheorem{theorem}{Theorem}

\begin{document}

\title{Entanglement Monogamy Relations of Qubit Systems}

\author{Xue-Na Zhu$^{1}$}
\author{Shao-Ming Fei$^{2,3}$}

\affiliation{$^1$School of Mathematics and Statistics Science, Ludong University, Yantai 264025, China\\
$^2$School of Mathematical Sciences, Capital Normal
University, Beijing 100048, China\\
$^3$Max-Planck-Institute for Mathematics in the Sciences, 04103 Leipzig, Germany}

\begin{abstract}
We investigate the monogamy relations related to the concurrence and the
entanglement of formation. General monogamy inequalities given by the $\alpha$th power
of concurrence and entanglement of formation are presented for $N$-qubit states.
The monogamy relation for entanglement of assistance is also established.
Based on these general monogamy relations, the residual entanglement of concurrence
and entanglement of formation are studied.
Some relations among the residual entanglement, entanglement of assistance and three tangle
are also presented.
\end{abstract}

\pacs{ 03.67.Mn,03.65.Ud}

\maketitle

\section{Introduction}

Quantum entanglement \cite{t1,t2,t3,t4,t5,t6} is an essential feature of quantum mechanics,
which distinguishes the quantum from classical world.
As one of the fundamental differences between quantum entanglement and classical
correlations, a key property of entanglement is that a quantum system entangled with one of other
systems limits its entanglement with the remaining others.
The monogamy relations give rise to the structures of entanglement in
the multipartite setting. Monogamy is also an essential feature
allowing for security in quantum key distribution \cite{k3}.

For a tripartite system $A$, $B$ and $C$,
the monogamy of an entanglement measure $\varepsilon$ implies that \cite{022309},
the entanglement between $A$ and $BC$ satisfies
$\varepsilon_{A|BC}\geq\varepsilon_{AB}+\varepsilon_{AC}$.
Such monogamy relations are not always satisfied by entanglement measures.
Although the concurrence $C$ and entanglement of formation $E$ do not satisfy
such monogamy inequality, it has been shown that the squared concurrence $C^2$ \cite{PRA80044301,C2}
and the squared entanglement of formation $E^2$ \cite{E1}
do satisfy the monogamy relations.

In this paper, we study the general monogamy inequalities satisfied by the $\alpha$th power
of concurrence $C^\alpha$ and the $\alpha$th power of entanglement of formation $E^\alpha$.
We show that $C^\alpha$ and $E^\alpha$ satisfy the monogamy inequalities
for $\alpha\geq 2$ and $\alpha\geq \sqrt{2}$, respectively.
The monogamy relations for the entanglement of assistance are also established.
Correspondingly, the residual entanglement of concurrence
and entanglement of formation are also investigated.

\section{Monogamy relation of concurrence}

For a bipartite pure state $|\psi\rangle_{AB}$ in vector space $H_A\otimes H_{B}$,
the concurrence is given by \cite{s7,s8,af}
\begin{equation}\label{CON}
C(|\psi\rangle_{AB})=\sqrt{2[1-Tr(\rho^2_A)]},
\end{equation}
where $\rho_A$ is reduced density matrix by tracing over the subsystem $B$,
$\rho_{A}=Tr_{B}(|\psi\rangle_{AB}\langle\psi|)$.
The concurrence for a tripartite mixed state $\rho_{AB}$ is defined by the
convex roof,
\begin{equation}\label{CONC}
C(\rho_{AB})=\min_{\{p_i,|\psi_i\rangle\}} \sum_i p_i C(|\psi_i\rangle),
\end{equation}
where the minimum (infimum) is taken over all possible decompositions of $\rho_{AB}=\sum_ip_i|\psi_i\rangle \langle \psi_i|$,
with $p_i\geq0$ and $\sum_ip_i=1$ and  $|\psi _{i}\rangle\in H_A\otimes H_{B}$.

For a tripartite state $|\psi\rangle_{ABC}$, the concurrence of assistance (CoA) is defined by \cite{s6,ca}
\begin{equation}
C_a(|\psi\rangle_{ABC})\equiv C_a(\rho_{AB})
=\max_{\{p_i,|\psi_i\rangle\}}\sum_ip_iC(|\psi_i\rangle),
\end{equation}
where the maximum (supremum) is taken over all possible decompositions of
$\rho_{AB}=Tr_{C}(|\psi\rangle_{ABC}\langle\psi|)=\sum_i p_i |\psi_i\rangle_{AB} \langle \psi_i|$.
When $\rho_{AB}=|\psi\rangle_{AB}\langle \psi|$ is a pure state, then one has
$C(|\psi\rangle_{AB})=C_{a}(\rho_{AB})$.

For an $N$-qubit state $|\psi\rangle_{AB_1...B_{N-1}}\in H_A\otimes H_{B_1}\otimes...\otimes H_{B_{N-1}}$, the concurrence
$C(|\psi\rangle_{A|B_1...B_{N-1}})$ of the state
$|\psi\rangle_{A|B_1...B_{N-1}}$, viewed as a bipartite
with partitions $A$ and $B_1B_2...B_{N-1}$, satisfies the Coffman-Kundu-Wootters (CKW)
inequality \cite{PRA80044301,C2},
\begin{equation}\label{cm}
C^2_{A|B_1B_2...B_{N-1}}\geq C^2_{AB_1}+C^2_{AB_2}+...+C^2_{AB_{N-1}},
\end{equation}
where $C_{AB_i}=C(\rho_{AB_i})$ is the concurrence of $\rho_{AB_i}=Tr_{B_1...B_{i-1}B_{i+1}...B_{N-1}}(\rho)$,
$C_{A|B_1B_2...B_{N-1}}=C(|\psi\rangle_{A|B_1...B_{N-1}})$.

Dual to the CKW inequality, the generalized monogamy relation
based on the concurrence of assistance was proved in Ref. \cite{dualmonogamy},
\begin{equation}\label{ca}
C^2(|\psi\rangle_{A|B_1...B_{N-1}})\leq \sum_{i=1}^{N-1}C^2_a(\rho_{AB_i}).
\end{equation}

The inequalities (\ref{cm}) and (\ref{ca}) are valid because, instead of the
concurrence and CoA, the squared concurrence and CoA are used.
In fact, besides the squared concurrence, one can get the following
general monogamy inequalities:

\begin{theorem}\label{TH1}
For any $2\otimes2...\otimes 2\otimes2$  mixed state $\rho\in H_A\otimes H_{B_1}\otimes...\otimes H_{B_{N-1}}$,
we have
\begin{equation}\label{a}
C^{\alpha}_{A|B_1B_2...B_{N-1}}\geq C^{\alpha}_{AB_1}+...+C^{\alpha}_{AB_{N-1}}
\end{equation}
for all $\alpha\geq 2$.
\end{theorem}

{\sf [Proof]}~ For arbitrary $2\otimes2\otimes2^{n-2}$ tripartite
state $\rho_{ABC}$, one has \cite{C2,xj},
 \begin{equation*}
 C^2_{A|BC}\geq C^2_{AB}+C^2_{AC}.
 \end{equation*}
If $\min\{C_{AB},C_{AC}\} =0$, obviously we have
$C^{\alpha}_{A|BC}\geq C^{\alpha}_{AB}+C^{\alpha}_{AC}.$
If $\min\{C_{AB},C_{AC}\} >0$, assuming $C_{AB}\geq C_{AC}$, we have
\begin{eqnarray}\label{CCC}
C^{\alpha}_{A|BC}
&\geq & (C^2_{AB}+C^2_{AC})^{\frac{\alpha}{2}}\\[2mm]\nonumber
&=&C^{\alpha}_{AB}
\Big(1+\frac{C^2_{AC}}{C^2_{AB}}\Big)^{\frac{\alpha}{2}}\\[2mm]\nonumber
&\geq&C^{\alpha}_{AB}
\left(1+\Big(\frac{C^2_{AC}}{C^2_{AB}}\Big)^{\frac{\alpha}{2}}\right)\\[2mm]\nonumber
&=&C^{\alpha}_{AB}+C^{\alpha}_{AC},
\end{eqnarray}
where the second inequality is due to the inequality $(1+x)^t\geq1+x^t$ for $x\leq1$ and $t\geq1$.

By partitioning the last qudit system C into two subsystems: a qubit system $C_1$
and a $2^{n-3}$-dimensional qudit system $C_2$, and using the above inequality repeatedly, one gets (\ref{a}).
\qed

\begin{theorem}\label{TH2}
For any $2\otimes2\otimes...\otimes2$ mixed state $\rho\in H_A\otimes H_{B_1}\otimes...\otimes H_{B_{N-1}}$ with
$C_{AB_i}\not=0$, $i=1,...,N-1$,
we have
\begin{equation}\label{t2}
C^{\alpha}_{A|B_{1}...B_{N-1}}
<C^{\alpha}_{AB_{1}}+...+C^{\alpha}_{AB_{N-1}}
\end{equation}
for $\alpha\leq0$.
\end{theorem}

{\sf [Proof]}~Similar to the proof of Theorem \ref{TH1}, we only need to prove the inequality is true
for arbitrary $2\otimes2\otimes2^{n-2}$ tripartite states,
\begin{eqnarray}
C^{\alpha}_{A|BC}
&\leq & (C^2_{AB}+C^2_{AC})^{\frac{\alpha}{2}}\\[2mm]\nonumber
&=&C^{\alpha}_{AB}
\Big(1+\frac{C^2_{AC}}{C^2_{AB}}\Big)^{\frac{\alpha}{2}}\\[2mm]\nonumber
&<&C^{\alpha}_{AB}
\left(1+\Big(\frac{C^2_{AC}}{C^2_{AB}}\Big)^{\frac{\alpha}{2}}\right)\\[2mm]\nonumber
&=&C^{\alpha}_{AB}+C^{\alpha}_{AC},
\end{eqnarray}
where the first  inequality is due to $\alpha\leq0$
and
the second inequality is due to $C^{\alpha}_{AB}>0$ and the inequality $(1+x)^t<1+x^t$ for $x>0$ and $t\leq0$.
\qed

In (\ref{t2}) we have assumed that all
$C_{AB_{i}}$, $i=1,...,N-1$, are nonzero. In fact, if one of them is zero,
the inequality still holds if one removes this term from the inequality.
Namely, if $C_{AB_i}=0$, then one has $C^{\alpha}_{A|B_1...B_{N-1}}
<C^{\alpha}_{AB_1}+...+C^{\alpha}_{AB_{i-1}}
+C^{\alpha}_{AB_{i+1}}+...+C^{\alpha}_{AB_{N-1}}$ for $\alpha\leq0$.

Theorem \ref{TH1} shows that the $\alpha$th power of concurrence $C^{\alpha}$
satisfies the monogamy inequality (\ref{a}) for $\alpha\geq 2$.
While Theorem \ref{TH2} shows that for $\alpha\leq 0$, the inequality is reversed.
However, for $0 < \alpha < 2$, the situation is not clear. Let us
consider the three-qubit case. Any three-qubit state $|\psi\rangle$ can be written in
the generalized Schmidt decomposition \cite{gx,xhg},
\begin{equation}\label{pure3}
\ba{rcl}
|\psi\rangle&=&\lambda_0|000\rangle+
\lambda_1e^{i\varphi}|100\rangle
+\lambda_2|101\rangle\\[2mm]
&&+\lambda_3|110\rangle+
\lambda_4|111\rangle,
\ea
\end{equation}
where $\lambda_i\geq0$, $i=0,...,4$, and $\sum_{i=0}^{4}\lambda_i^2=1$.
From  Eq.(\ref{CON}) and Eq.(\ref{CONC}), we have
$C_{A|BC}=2\lambda_0\sqrt{\lambda^2_2+\lambda^2_3+\lambda^2_4},$
$C_{AB}=2\lambda_0\lambda_2,$ and $C_{AC}=2\lambda_0\lambda_3.$
Without loss of generality, we set $\lambda_0=\cos\theta_0,$
$\lambda_1= \sin\theta_0\cos\theta_1,$ $\lambda_2=
\sin\theta_0\sin\theta_1\cos\theta_2,$ $\lambda_3=
\sin\theta_0\sin\theta_1\sin\theta_2\cos\theta_3, $ and $\lambda_4=
\sin\theta_0\sin\theta_1\sin\theta_2\sin\theta_3$,
$\theta_i\in[0,\frac{\pi}{2}]$. Then we have
\begin{equation}\label{tong}
\begin{aligned}
&C^{\alpha}_{A|BC}-C^{\alpha}_{AB}-C^{\alpha}_{AC}\\
&=(2\lambda_0)^{\alpha}
\Big[(\lambda^2_2+\lambda^2_3+\lambda^2_4)^{\frac{\alpha}{2}}-\lambda^{\alpha}_2
-\lambda^{\alpha}_3\Big]\\
&=(2\lambda_0)^{\alpha}\sin^{\alpha}\theta_0\sin^{\alpha}\theta_1
\Big[
1-\cos^{\alpha}\theta_2-\sin^{\alpha}\theta_2\cos^{\alpha}\theta_3
\Big].
\end{aligned}
\end{equation}
From (\ref{tong}) we have
$C^{\alpha}_{A|BC}\geq C^{\alpha}_{AB}+C^{\alpha}_{AC}$ for $\alpha\geq2$.
While for $\alpha\leq0$ one has $C^{\alpha}_{A|BC}<C^{\alpha}_{AB}+C^{\alpha}_{AC}$.
However, for $0<\alpha<2$,
one can see that the sign of $(C^{\alpha}_{A|BC}- C^{\alpha}_{AB}-C^{\alpha}_{AC})$ is not certain.

\section{Residual entanglement of concurrence}
Similar to the three tangle of concurrence,
for the three qubit state $|\psi\rangle_{ABC}\in H_A\otimes H_B\otimes H_C$, we can define the residual entanglement
\begin{equation}\label{tauc}
\tau^{C}_{\alpha}(|\psi\rangle_{ABC})
=C^{\alpha}_{A|BC}-C^{\alpha}_{AB}-C^{\alpha}_{AC},
\end{equation}
where $\alpha\geq2.$

\begin{theorem}\label{TH3}
For any three qubit pure state $|\psi\rangle\in H_A\otimes H_B\otimes H_C$,

(1) $|\psi\rangle$ is bipartite separable state if and only if
for any $\alpha\geq2$,
\begin{equation*}
\tau^{C}_{\alpha}(|\psi\rangle)=0;
\end{equation*}

(2) $|\psi\rangle$ is genuine entangled if and only if
there is an $\alpha\geq2$ such that
\begin{equation*}
\tau^{C}_{\alpha}(|\psi\rangle)>0.
\end{equation*}
\end{theorem}

{\sf [Proof]}~
(1) If $|\psi\rangle$ is  bipartite separable state,
without loss of generality, we assume that $|\psi\rangle$ is a $B|AC$ bipartite separable state,
then we have $C_{B|AC}=C_{BA}=C_{BC}=0$. From (\ref{pure3}) we have
$\lambda_0=0$  and $|\lambda_1\lambda_4e^{i\varphi}-\lambda_2\lambda_3|=0$, or
$\lambda_3=\lambda_4=0$ and $|\lambda_1\lambda_4e^{i\varphi}-\lambda_2\lambda_3|=0.$ For both the
above two cases, we have $C_{A|BC}=C_{AC}$ and $C_{AB}=0$. Therefore
$\tau^{C}_{\alpha}(|\psi\rangle)=0$ for all $\alpha$.

If $\tau_{\alpha}^C(|\psi\rangle)=0$
for any $\alpha\geq2$, then we obtain
$C^{2\alpha}_{A|BC}-C^{2\alpha}_{AB}-C^{2\alpha}_{AC}
=(C^{\alpha}_{A|BC})^2-C^{2\alpha}_{AB}-C^{2\alpha}_{AC}
=(C^{\alpha}_{AB}+C^{\alpha}_{AC})^2
-C^{2\alpha}_{AB}-C^{2\alpha}_{AC}
=2C^{\alpha}_{AB}C^{\alpha}_{AC}=0$,
i.e. either $C^{\alpha}_{AB}$ or $C^{\alpha}_{AC}$ is zero.
Without loss of generality, assuming $C_{AB}=0$, we have $\lambda_0=0$ or $\lambda_3=0$.
If $\lambda_0=0$, then $C_{A|BC}=0$. Hence $|\psi\rangle$ is $A|BC$ a bipartite separable state.
If $\lambda_0\not=0$ and $\lambda_3=0$, since $\tau^C_2=0$, we have $\lambda_4=0$. Hence $C_{B|AC}=0$,
i.e. $|\psi\rangle$ is $B|AC$ bipartite separable.

(2) By using the above proof and the fact that $\tau_{\alpha}^C(|\psi\rangle)\geq0$ for all $\alpha\geq2$ from Theorem \ref{TH1},
one gets the result directly.
\qed

\begin{figure}
  \centering
  \includegraphics[width=6.5cm]{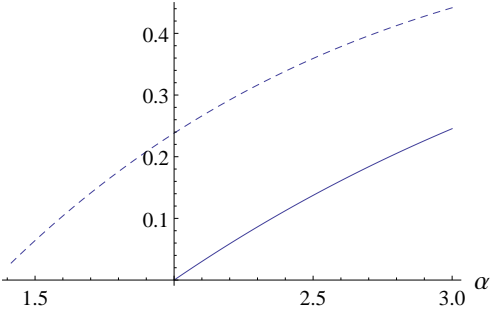}
  \caption{Solid line: $\tau^{C}_{\alpha}(|W\rangle)$ as a function of $\alpha$ ($\alpha\geq2$); dashed
  line: $\tau^{E}_{\alpha}(|W\rangle)$ as a function of $\alpha$ ($\alpha\geq\sqrt{2}$).}
  \label{t}
\end{figure}

{\it Example 1.} Let us consider the $W$-state
\be\label{w}
|W\rangle=\frac{1}{\sqrt{3}}(|100\rangle+|010\rangle+|001\rangle).
\ee
We have $\tau^{C}_{\alpha}(|W\rangle)=(\frac{2}{\sqrt{3}})^{\alpha}
\Big(\sqrt{(\frac{2}{3})^{\alpha}}-2(\frac{1}{\sqrt{3}})^{\alpha}\Big).$
For $\alpha=2$, $\tau^{C}_{\alpha}$ is just the three tangle of
concurrence. As $\tau^C_{2}(|W\rangle)=0$, the three tangle of
concurrence can not capture the genuine entanglement of the W-state.
Nevertheless, for $\alpha>2$, our residual entanglement of concurrence
$\tau^{C}_{\alpha}(|W\rangle)>0$, see Fig. (\ref{t}).

\section{Monogamy inequality for EoF}

The entanglement of formation (EoF) \cite{C. H. Bennett,D. P. DiVincenzo} is
a well-defined important measure of entanglement for bipartite systems.
Let $H_A$ and $H_B$ be $m$- and $n$-dimensional $(m\leq n)$ vector spaces, respectively.
The EoF of a pure state $|\psi\rangle\in H_A\otimes H_B$ is defined by
\begin{equation}\label{S}
E(|\psi\rangle)=S(\rho_A),
\end{equation}
where $\rho_A=Tr_{B}(|\psi\rangle\langle\psi|)$
and $S(\rho)=-Tr(\rho\log_2\rho)$. For a bipartite
mixed state $\rho_{AB}\in H_A\otimes H_B$, the entanglement of formation is given by
\begin{equation}\label{EE}
E(\rho_{AB})=\min_{\{p_i,|\psi_i\rangle\}}\sum_ip_iE(|\psi_i\rangle),
\end{equation}
with the minimum (infimum) taking over all possible decompositions of
$\rho_{AB}$ in a mixture of pure states
$\rho_{AB}=\sum_ip_i|\psi_i\rangle \langle \psi_i|$, where
$p_i\geq0$ and $\sum_ip_i=1$. The corresponding
entanglement of assistance (EoA) \cite{OC} is defined
in terms of the entropy of entanglement \cite{Ea} for a tripartite pure
state $|\psi\rangle_{ABC}$,
\begin{equation}
E_a(|\psi\rangle_{ABC})\equiv E_a(\rho_{AB})=\max_{\{p_i,|\psi_i\rangle\}}\sum_ip_iE(|\psi_i\rangle),
\end{equation}
where the maximum (supremum) is taken over all possible decompositions of $\rho_{AB}=Tr_{C}(|\psi\rangle_{ABC})=\sum_ip_i|\psi_i\rangle \langle \psi_i|$,
with $p_i\geq0$ and $\sum_ip_i=1$.

Denote $f(x)=H\left(\frac{1+\sqrt{1-x}}{2}\right),$
where $H(x)=-x\log_2(x)-(1-x)\log_2(1-x).$
From Eq. (\ref{S}) and Eq. (\ref{EE}), one has
$E(|\varphi\rangle)=f\Big(C^2(|\varphi\rangle)\Big)$ for $2\otimes m$ $(m\geq2)$ pure state $|\varphi\rangle$, and
$E(\rho)=f\Big(C^2(\rho)\Big)$ for two qubit mixed state $\rho$ \cite{Eof}.
It is obviously that $f(x)$ is a  monotonically increasing  function for $0\leq x\leq1$.
$f(x)$ satisfies the following relations:
\begin{equation}\label{fx}
f^{\sqrt{2}}(x^2+y^2)\geq f^{\sqrt{2}}(x^2)+f^{\sqrt{2}}(y^2),
\end{equation}
\begin{equation}\label{fx2}
f(x^2+y^2)\leq f(x^2)+f(y^2),
\end{equation}
where $f^{\sqrt{2}}(x^2+y^2)=(f(x^2+y^2))^{\sqrt{2}}$.

It has been shown that the entanglement of formation does not satisfy
the inequality $E_{AB}+E_{AC}\leq E_{A|BC}$ \cite{E2}.
In \cite{oneof} the authors showed that EoF is a monotonic function
$E^2(C^2_{A|B_1B_2...B_{N-1}})\geq E^2(\sum_{i=1}^{N-1}C^2_{AB_i})$.
It is further proved that for $N$-qubit systems, one has \cite{E1},
$$
E_{A|B_1B_2...B_{N-1}}^2\geq E_{AB_1}^2+E_{AB_2}^2+...+E_{AB_{N-1}}^2.
$$
In fact, generally we can prove the following results:

\begin{theorem}\label{TH4}
For any $N$-qubit mixed state $\rho\in H_{A}\otimes H_{B_1}\otimes...\otimes H_{B_{N-1}}$,
the entanglement of formation $E(\rho)$ satisfies
\begin{equation}\label{th4}
E_{A|B_1B_2...B_{N-1}}^\alpha\geq E_{AB_1}^\alpha+E_{AB_2}^\alpha
+...+E_{AB_{N-1}}^\alpha,
\end{equation}
where $\alpha\geq\sqrt{2}$, $E_{A|B_1B_2...B_{N-1}}$
is the entanglement of formation   of  $\rho$  in bipartite
partition $A|B_1B_2...B_{N-1}$, and $E_{AB_i}$, $i=1,2...,N-1$, is the
entanglement of formation of the mixed states $\rho_{AB_i}
=Tr_{B_1B_2...B_{i-1}B_{i+1}...B_{N-1}}(\rho)$.
\end{theorem}

{\sf [Proof]}~Denote $t={\alpha}/{\sqrt{2}}$.
For $\alpha\geq\sqrt{2}$, we have
\begin{equation}\label{d}
\ba{rcl}
f^{\alpha}(x^2+y^2)&=&(f^{\sqrt{2}}(x^2+y^2))^t\\[2mm]
&\geq&(f^{\sqrt{2}}(x^2)+f^{\sqrt{2}}(y^2))^t\\[2mm]
&\geq&(f^{\sqrt{2}}(x^2))^t+(f^{\sqrt{2}}(y^2))^t\\[2mm]
&=&f^{\alpha}(x^2)+f^{\alpha}(y^2),
\ea
\end{equation}
where the first inequality is due to the inequality (\ref{fx}),
and the second inequality is obtained from a similar consideration in the  proof of the second
inequality in (\ref{CCC}).

Let $\rho=\sum_ip_i|\psi_i\rangle\langle\psi_i|\in H_A
\otimes H_{B_1}\otimes H_{B_2}\otimes...\otimes H_{B_{N-1}}$ be the optimal
decomposition of $E_{A|B_1B_2...B_{N-1}}(\rho)$ for the $N$-qubit mixed state $\rho$,
we have
\begin{equation*}
\ba{l}
E_{A|B_1B_2...B_{N-1}}(\rho)\\[3mm]
= \sum_ip_iE_{A|B_1B_2...B_{N-1}}(|\psi_i\rangle)\\[3mm]
= \sum_ip_if\left(C^2_{A|B_1B_2...B_{N-1}}(|\psi_i\rangle)\right)\\[3mm]
\geq f\Big(\Big[\sum_ip_iC_{A|B_1B_2...B_{N-1}}(|\psi_i\rangle)\Big]^2\Big)\\[3mm]
\geq f(C^2_{A|B_1B_2...B_{N-1}}(\rho))\\[3mm]
\geq f(C^2_{AB_1}+C^2_{AB_2}+...+C^2_{AB_{N-1}})\\[3mm]
\geq \sqrt[\alpha]{f^\alpha(C^2_{AB_1})
+f^\alpha(C^2_{AB_2})+...+f^\alpha(C^2_{AB_{N-1}})}\\[3mm]
= \sqrt[\alpha]{E_{AB_1}^\alpha +E_{AB_2}^\alpha+...
+E_{AB_{N-1}}^\alpha},
\ea
\end{equation*}
where
the first inequality is due to that $f(x^2)$ is a convex function of $x$. 
Due to the definition of concurrence (\ref{CONC}) and that $f(x)$ is a monotonically increasing function, we obtain
the second inequality.
We have used the monogamy inequality (\ref{cm}) for $N$-qubit states $\rho$ to obtain the third inequality. The last
inequality is due to the inequality (\ref{d}). Since for any $2\otimes 2$ quantum
state $\rho_{AB_i}$, $E(\rho_{AB_i})$ satisfies
$E(\rho_{AB_i})=H\left(\frac{1+\sqrt{1-C^2(\rho_{AB_i})}}{2}\right)
=f(C^2(\rho_{AB_i})),$ one gets the last equality. \qed

The inequality (\ref{th4}) in Theorem \ref{TH4} shows
that the $\alpha$th power of EoF satisfies
the monogamy inequality for any $\alpha\geq\sqrt{2}$, which is a little different
from the case of concurrence in which $\alpha\geq {2}$.
As for the entanglement of assistance, we have the following conclusion:

\begin{theorem}\label{TH5}
For any $N$-qubit pure state $|\psi\rangle\in H_A
\otimes H_{B_1}\otimes...\otimes H_{B_{N-1}}$, the entanglement of assistance
satisfies
\begin{equation}
E(|\psi\rangle_{A|B_1B_2...B_{N-1}})\leq
\sum_{i=1}^{N-1}E_a(\rho_{AB_i}),
\end{equation}
where $E(|\psi\rangle_{A|B_1B_2...B_{N-1}})$ is  the entanglement of formation   of  $|\psi\rangle$  in bipartite
partition $A|B_1B_2...B_{N-1}$, and  $\rho_{AB_i}
=Tr_{B_1...B_{i-1}B_{i+1}...B_{N-1}}(|\psi\rangle\langle\psi|)$.
\end{theorem}

{\sf [Proof]}~Let $\rho_{AB}=\sum_ip_i|\psi_i\rangle\langle\psi_i|$ be the optimal
decomposition of $C_a(\rho_{AB})$. We have
\begin{equation}\label{Ea}
\ba{rcl}
E_a(\rho_{AB})&\geq&\sum_ip_iE(|\psi_i\rangle)\\[2mm]
&=&\sum_ip_if\Big(C_a^2(|\psi_i\rangle)\Big)\\[2mm]
&\geq&f\left(\Big[\sum_ip_iC_a(|\psi_i\rangle)\Big]^2\right)\\[4mm]
&=&f(C^2_{a}(\rho_{AB})),
\ea
\end{equation}
where the first equality is due to that,
for a pure state, one has  $\rho_{AB}=\rho_{AB}^2$ and $C(\rho_{AB})=C_a(\rho_{AB})$ \cite{dualmonogamy}.
The second inequality is due to that $f(x^2)$ is a convex function of $x$. 

Therefore for an $N$-qubit pure state $|\psi\rangle_{AB_1...B_{N-1}}$, we have
\begin{equation*}
\ba{l}
E(|\psi\rangle_{A|B_1...B_{N-1}})=f(C^2(|\psi\rangle_{A|B_1...B_{N-1}}))\\[2mm]
~~~~~~~\leq f\Big(C^2_{a}(\rho_{AB_1})+... +C^2_{a}(\rho_{AB_{N-1}})\Big)\\[2mm]
~~~~~~~\leq f\Big(C^2_{a}(\rho_{AB_1})\Big)+...+f\Big(C^2_{a}(\rho_{AB_{N-1}})\Big)\\[2mm]
~~~~~~~\leq E_{a}(\rho_{AB_1})+...+E_{a}(\rho_{AB_{N-1}}).
\ea
\end{equation*}
The first inequality is due to
Eq. (\ref{ca}). We have used the inequality
(\ref{fx2}) to get the second inequality. The last inequality is due to (\ref{Ea}).
 \qed

\section{Residual entanglement of EoF}

Similar to the residual entanglement (\ref{tauc}) defined by the
$\alpha$th ($\alpha\geq 2$) power of concurrence,
we can define the residual entanglement by the
$\alpha$th ($\alpha\geq \sqrt{2}$) power of EoF
for a three qubit pure state $|\psi\rangle_{ABC}$:
\begin{equation}
\tau^E_{\alpha}(|\psi\rangle_{ABC})=E^\alpha_{A|BC}-E^\alpha_{AB}-E^\alpha_{AC}\geq 0.
\end{equation}

As an example, let us consider again the W-state (\ref{w}).
We have $E_{AB}=E_{AC}=0.550048$ and $E_{A|BC}=0.918296$.
Therefore $\tau^E_{\alpha}=0.918296^\alpha-2(0.550048)^\alpha$, see Fig. (\ref{t}).

Here it should be noted that, different from the
residual entanglement of concurrence, the residual entanglement of EoF
depends on which qubit is chosen to be $A$.

In the following we give some relations among the residual entanglement of EoF, entanglement of assistance
and three tangle.
\begin{theorem}\label{TH6}
For a three qubit pure state $|\psi\rangle_{ABC}$, we have
\begin{equation}
\tau^E_{\alpha}(|\psi\rangle_{ABC})\geq f^2\left(\tau^C_2(|\psi\rangle_{ABC})\right),
\end{equation}
and
\begin{equation}
 E_a^{\alpha}(\rho_{AB})\geq E^\alpha(\rho_{AB})+f^\alpha(\tau^C_2(|\psi\rangle_{ABC})),
\end{equation}
where $\alpha\geq\sqrt{2}$, $\rho_{AB}=Tr_C(|\psi\rangle_{ABC}\langle\psi|)$ and
$\tau^C_2(|\psi\rangle_{ABC})$ is the three tangle of concurrence.
\end{theorem}

{\sf [Proof]}~ According to the definition of
$\tau^E_{\alpha}(|\psi\rangle_{ABC})$, we have
\begin{equation*}
\ba{rcl}
\tau^E_{\alpha}(|\psi\rangle_{ABC})
&=&E^{\alpha}_{A|BC}-E^{\alpha}_{AB}-E^{\alpha}_{AC}\\[2mm]
&=&f^{\alpha}(C^2_{A|BC})-f^{\alpha}(C^2_{AB})-f^{\alpha}(C^2_{AC})\\[2mm]
&=&f^{\alpha}(C^2_{AB}+C^2_{AC}+\tau_2^C(|\psi\rangle_{ABC}))\\[2mm]
&&-f^{\alpha}(C^2_{AB})-f^{\alpha}(C^2_{AC})\\[2mm]
&\geq&f^{\alpha}\left(\tau_2^C(|\psi\rangle_{ABC})\right),
\ea
\end{equation*}
where the third equality is due to the definition of the three tangle $\tau_2^C$.
We have used the Eq. (\ref{d}) to get the last inequality.

Accounting to that for a $2\otimes2\otimes m$ quantum pure state $|\psi\rangle_{ABC}$,
$C^2_{a}(\rho_{AB})= C^2(\rho_{AB})+\tau^C_2(|\psi\rangle_{ABC})$ \cite{022324}, we have
\begin{equation*}
\ba{rcl}
E_a(\rho_{AB})
&\geq&f(C^2_{a}(\rho_{AB}))\\[3mm]
&=&f(C^2(\rho_{AB})+\tau^C_2(|\psi\rangle_{ABC}))\\[3mm]
&\geq&\sqrt[\alpha]{f^\alpha(C^2(\rho_{AB}))
+f^\alpha(\tau^C_2(|\psi\rangle_{ABC}))}\\[3mm]
&=&\sqrt[\alpha]{E^\alpha(\rho_{AB})
+f^\alpha(\tau^C_2(|\psi\rangle_{ABC}))},
\ea
\end{equation*}
where we have used the inequality (\ref{Ea}) to obtain the first
inequality and the Eq. (\ref{fx}) to get the last inequality.
\qed

The relations among entanglement of formation, entanglement of assistance
and three tangle given in Theorem \ref{TH6} can be
used to obtain a lower bound of EoA. Let us consider the
following example.

\begin{figure}
  \centering
  \includegraphics[width=6.5cm]{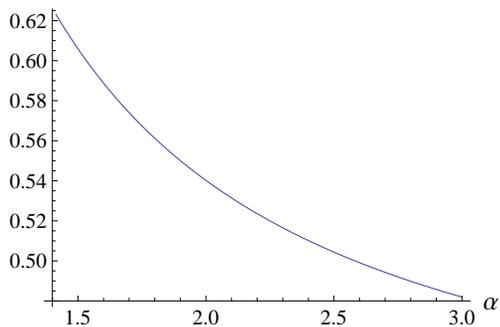}
  \caption{The lower bound of $E_{a}(\rho_{AB})$ for state (\ref{fig2}) with $\alpha\geq\sqrt{2}$.}
  \label{3}
\end{figure}

{\it Example 2.}
Superpositions of the Greenberger-Horne-Zeilinger
(GHZ)-state and the W-state (\ref{w}):
\be\label{fig2}
|\Psi\rangle_{ABC}=\sqrt{\frac{1}{2}}|GHZ\rangle-\sqrt{\frac{1}{2}}|W\rangle,
\ee
where $|GHZ\rangle=\frac{1}{\sqrt{2}}(|000\rangle+|111\rangle)$.
According to Theorem 6 we obtain the lower bound of $E_{a}(\rho_{AB})$,
$E_a(\rho_{AB})\geq (E^\alpha(\rho_{AB})+f^\alpha(\tau^C_2(|\psi\rangle_{ABC})))^{\frac{1}{\alpha}}$,
where $\rho_{AB}=Tr_{C}(|\Psi\rangle_{ABC}\langle\Psi|)$, see Fig. (\ref{3}).
From Fig. (\ref{3}), one gets that the optimal lower bound of $E_{a}(\rho_{AB})$ is $0.623$ at $\alpha=\sqrt{2}$.

\section{Conclusion}

Entanglement monogamy is a fundamental property of multipartite
entangled states. We have investigated the monogamy relations
related to the concurrence and the
entanglement of formation generally for $N$-qubit states.
We also proved that the
entanglement of assistance satisfies the monogamy inequality
$E(|\psi\rangle_{A|B_1B_2...B_{N-1}})\leq
\sum_{i=1}^{N-1}E_{a}(\rho_{AB_i})$. To study the genuine tripartite entanglement,
we investigated the residual entanglement of concurrence $\tau^C_{\alpha}(|\psi\rangle_{ABC})$
and the residual entanglement of entanglement of formation
$\tau^E_{\alpha}(|\psi\rangle_{ABC})$.
By exploring the relations among the residual entanglement, entanglement of assistance and three tangle,
we have presented a bound of $E_a(\rho)$.
Our approach may be used to study further the monogamy properties
related to other quantum entanglement measures such as negativity and to
quantum correlations such as quantum discord.

\bigskip
\noindent{\bf Acknowledgments}\, \,
This work is supported by the NSFC under number 11275131.

\end{document}